\documentclass{elsart}
\usepackage{epsfig}
\usepackage{amssymb}
\journal{Nuclear Physics A}
\begin{document}

\begin{frontmatter}

\title{DWBA analysis of the $^{13}$C($^6$Li,$d$)$^{17}$O
reaction at 10 MeV/nucleon and its astrophysical implications}
\author[FSU]{N. Keeley}\footnote{Corresponding author,
E-mail: keeley@nucmar.physics.fsu.edu},
\author[FSU]{K.W. Kemper},
\author[VAEC]{Dao T. Khoa}
\address[FSU]{Department of Physics, Florida State University, Tallahassee, Florida 32306-4350 USA.}
\address[VAEC]{Institute for Nuclear Science and
 Technique, VAEC, P.O. Box 5T-160, Nghia Do, Hanoi, Vietnam.}

\begin{abstract}
The value of the $\alpha$ spectroscopic factor ($S_\alpha$) of the 6.356 MeV $1/2^+$ state of
$^{17}$O is believed to have significant astrophysical implications due to the
importance of the $^{13}$C($\alpha$,$n$)$^{16}$O reaction as a possible source of neutron
production for the $s$ process. To further study this effect, an accurate
measurement of the $^{13}$C($^6$Li,$d$)$^{17}$O reaction at $E_{\rm lab}=60$ MeV
has been performed recently by Kubono {\sl et al.\/}, who found a new value
for the spectroscopic factor of the 6.356 MeV $1/2^+$ state of $^{17}$O based on
a distorted wave Born approximation (DWBA) analysis of these data. This new value,
$S_\alpha\approx 0.011$, is surprisingly much smaller than those used previously
in astrophysical calculations ($S_\alpha\approx 0.3-0.7$) and thus poses a
serious question as to the role of the $^{13}$C($\alpha$,$n$)$^{16}$O reaction as a
source of neutron production. In this work we perform a detailed analysis of the
same $^{13}$C($^6$Li,$d$)$^{17}$O data within the DWBA as well as the coupled
reaction channel (CRC) formalism. Our analysis yields an $S_\alpha$ value of over an order of
magnitude larger than that of Kubono {\sl et al.\/} for the 6.356 MeV $1/2^+$ state of $^{17}$O.
\end{abstract}

\end{frontmatter}

\section{Introduction}
The slow neutron capture, or $s$ process is thought to be the production
mechanism for approximately half of all heavy elements in the universe
\cite{kubono,aoki}. The asymptotic giant branch (AGB) phase of low and
intermediate mass stars is thought to be the most likely astrophysical site for
this process \cite{aoki}. Although the neutron source for the $s$ process has
still not been well identified, the most recent studies seem to prefer the
$^{13}$C($\alpha$,$n$)$^{16}$O reaction as the main neutron source in AGB
stars at low temperatures \cite{aoki,gallino}.

Models of the $s$ process depend critically on the neutron flux produced by the
$^{13}$C($\alpha$,$n$)$^{16}$O reaction. However, the astrophysical $S$ factor
for this reaction, defined by:
\begin{equation}
S(E) = \sigma(E)E \exp{(2\pi\eta)} \label{e1}
\end{equation}
where $\sigma(E)$ is the cross section and $\eta$ the Coulomb parameter, has
been determined experimentally down to a centre of mass energy of only 270 keV
\cite{drotleff}, whereas the reaction takes place predominantly at energies
below this point.

As the cross section for the $^{13}$C($\alpha$,$n$)$^{16}$O reaction decreases
extremely rapidly as the incident $\alpha$ energy gets lower, direct measurement
of the reaction rate at lower energies is very difficult. Extrapolations
\cite{drotleff,hale} suggest a rapid increase of the $S$ factor as the incident
$\alpha$ energy reduces to zero, which has been ascribed to the effect of
resonances in $^{17}$O \cite{hale}, the $J^\pi$ = $1/2^-$ level at 5.94 MeV and
the $1/2^+$ level at 6.356 MeV, both very close to the $\alpha$ + $^{13}$C
threshold of 6.359 MeV. However, due to the experimental uncertainties, the data are also
consistent with a constant, horizontal extrapolated $S$ factor \cite{drotleff}.

In a recent paper, Kubono {\sl et al.\/} \cite{kubono} suggested that a better way
to determine the $S$ factor at lower energies is via a
direct $\alpha$ transfer measurement. The reaction chosen for this study was the
$^{13}$C($^6$Li,$d$)$^{17}$O transfer at $E_{\rm lab}=60$ MeV and 
angular distributions for transfers leading to the 0.0 MeV $5/2^+$, 0.87 MeV
$1/2^+$, 3.055 MeV $1/2^-$, 3.84 MeV $5/2^-$, 4.55 MeV $3/2^-$ and 6.356 MeV
$1/2^+$ states of $^{17}$O were accurately measured. The $\alpha$ spectroscopic
factor $S_\alpha$ of each state was determined from a finite-range
distorted-wave Born approximation (DWBA) analysis \cite{kubono} of these
data. Using the value of $S_\alpha$ obtained for the 6.356 MeV $1/2^+$
state, its $\alpha$ width was determined and the cross section of the
$^{13}$C($\alpha$,$n$)$^{16}$O reaction through the tail of this sub-threshold
resonance calculated. The astrophysical $S$ factor was then calculated using
Eq.~(\ref{e1}).

The DWBA analysis of Kubono {\sl et al.\/} gave a very small value of
$S_\alpha\approx 0.011$ for the 6.356 MeV $1/2^+$ state of $^{17}$O when
normalised by the same scaling factor used to obtain $S_\alpha=0.25$ for the 3.055
MeV $1/2^-$ state. This leads to a very small contribution of the $1/2^+_2$ sub-threshold
state to the reaction rate at low energies and thus to an astrophysical $S$
factor that is essentially constant below $E_{\rm c.m.}=300$ keV. Since such a
result poses serious questions about the mechanism of neutron production for the
$s$ process, a reanalysis of the Kubono {\sl et al.\/} data was undertaken.
In the present work we present the results of detailed DWBA and coupled reaction
channels (CRC) analyses of these same data to determine whether $S_\alpha$
for the 6.356 MeV $1/2^+$ state of $^{17}$O is actually so small. Our results
show that the original DWBA analysis of Kubono {\sl et al.\/} led to incorrect 
conclusions, and that their data 
are compatible with a much larger value of $S_\alpha$ for the 6.356 MeV
$1/2^+$ state. The present analysis is consistent with previous extrapolations of the
astrophysical $S$ factor for the $^{13}$C($\alpha$,$n$)$^{16}$O reaction that
indicate a rapid increase as the incident $\alpha$ energy approaches zero.

\section{The DWBA calculations}

Initially, we carried out a DWBA analysis as similar as possible to that of Kubono {\sl et al.\/} \cite{kubono},
but using the code FRESCO \cite{thompson}. We employed the same $\alpha$ + $^{13}$C  binding potential and
the same $^6$Li + $^{13}$C and $d$ + $^{17}$O optical potentials as in the original analysis \cite{kubono}.
However, all calculations presented here were carried out including the full complex remnant term unless otherwise
stated. The remnant term occurs in the residual interaction $W$ in the expression for the transition amplitude
in the DWBA. For the reaction $A + a(=b+x) \rightarrow B(=A+x) + b$ the residual interaction is 
defined as follows for post form DWBA:
\begin{equation}
W_\beta = V_{bB} - U_\beta = V_{bx} + (V_{bA} - U_\beta)
\end{equation}
and in the prior form:
\begin{equation}
W_\alpha = V_{aA} - U_\alpha = V_{xA} + (V_{bA} - U_\alpha).
\end{equation} 
The remnant terms are the quantities in parentheses, $U_\alpha$ and $U_\beta$ are the 
(complex) optical model potentials in the entrance and exit channels, respectively and $V_{bx}$ and
$V_{xA}$ are the potentials binding the transferred particle ($x$) to the projectile and target cores,
respectively. The quantity $V_{bA}$ is a (complex) optical potential operating between the projectile
and target cores. See Satchler \cite{satchler} for a full discussion of these terms.  

The use of the full
remnant term enables good agreement between post and prior formulations of the DWBA to be obtained, which is
not possible when no remnant term is included. We show results for the post formulation, although we did
perform test calculations using the prior form to ensure that agreement was obtained. As a further check
we compared calculations performed using FRESCO with no remnant term (using the post formulation)
with calculations carried out with the code DWUCK5 \cite{kunz}. For each transfer considered here excellent
agreement between the two codes was obtained.

\begin{figure}
\begin{center}
\psfig{figure=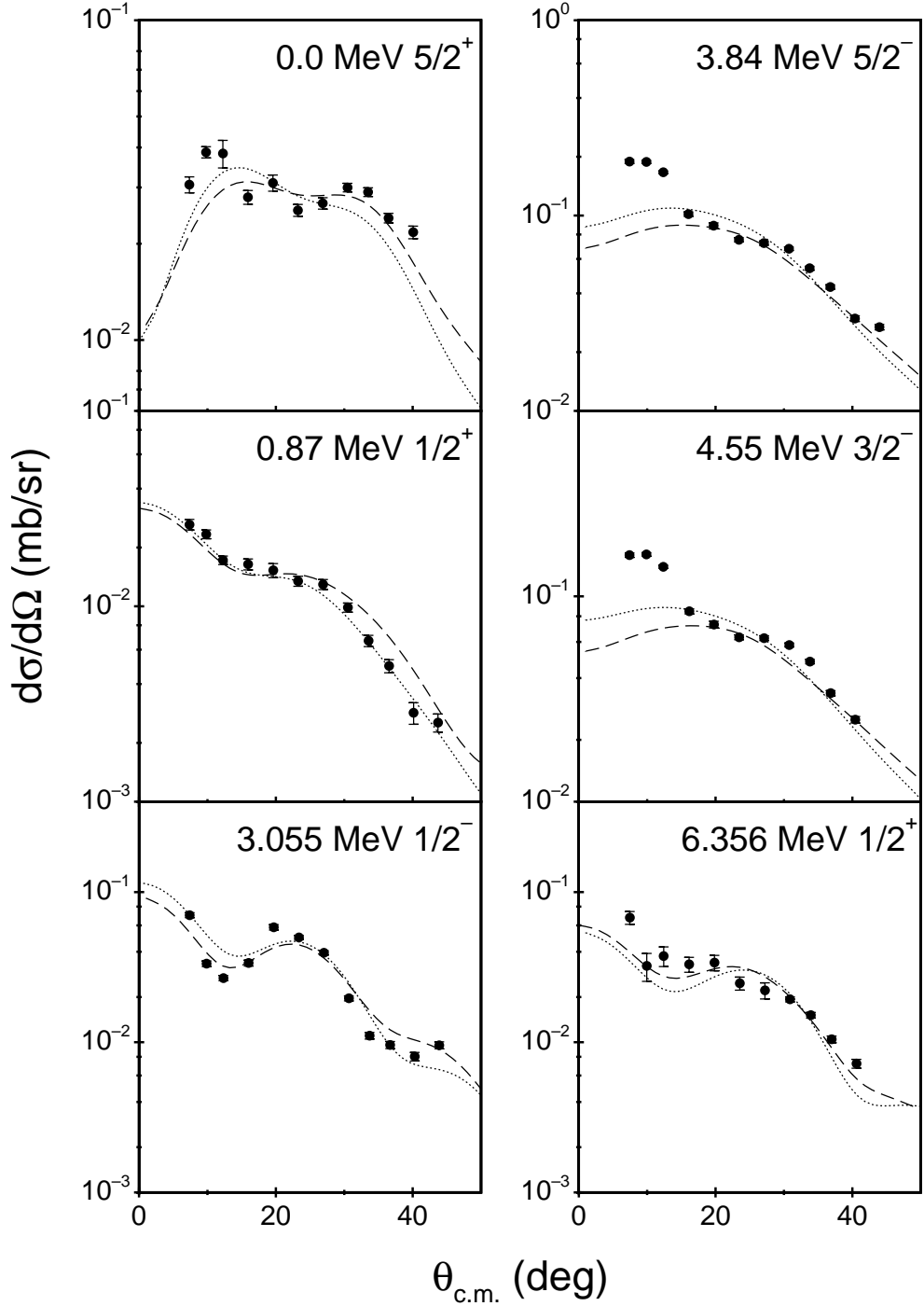,clip=,width=13.0cm}
\end{center}
\caption{\label{f1} DWBA calculations compared to the data. The dashed curves denote
calculations including the full complex remnant term, while the dotted curves
denote calculations with no remnant term. All calculations use the post form of
the DWBA.}
\end{figure}

The entrance channel $^6$Li + $^{13}$C optical potential was set 1 of Kubono
{\sl et al.\/} \cite{kubono} and we used their set 3 for the exit channel $d$ +
$^{17}$O potential \cite{cooper}. The $\alpha$ + $d$ binding potential was that
of Kubo and Hirata \cite{kubo}, with the $\alpha$ particle assumed to be in a
relative $2s$ state with respect to the $d$ core. Throughout we adopt the
convention that the number of radial nodes includes that at the origin but
excludes that at infinity. The $\alpha$ + $^{13}$C binding potential had a
radius of $1.25 \times (4^{1/3} + 13^{1/3})$ fm and diffuseness 0.65 fm, the
depth being adjusted to give the correct binding energy for each $^{17}$O state
considered. The number of nodes $N$ and orbital angular momentum of the
transferred $\alpha$ particle with respect to the $^{13}$C core $L$ were fixed
by the oscillatory energy conservation relation $2(N-1)+L=\sum_{i=1}^4
2(n_i-1)+l_i$, where $(n_i,l_i)$ are the corresponding single-nucleon shell
quantum numbers. The values of $N$ and $L$ used and the spectroscopic factors 
$S_\alpha$ obtained in the analysis are
given in Table~\ref{t1}. The spectroscopic factor for the $\alpha$+$d$ overlap
was taken to be 1.0. Note that the $S_\alpha$ values implicitly contain the
$C^2$ term, where $C$ is the isospin Clebsch-Gordan coefficient; however, in
this case $C = 1$. As can be seen from Table~\ref{t1} the contribution from the
remnant term in the DWBA calculations is substantial and accounts for more
than 60\% of the $S_\alpha$ strength for each $^{17}$O state. The results of the
DWBA calculations are compared with the data in Fig.~\ref{f1}.
\begin{table}\small
\caption{\small Values of $N$, $L$ and $S_\alpha$ used in the DWBA calculations.
Values of $S_\alpha$ are given for calculations with the full complex remnant
term and calculations with no remnant term included.} \label{t1}
\begin{tabular}{c c c c c c} \hline
State & $J^\pi$ & $N$ & $L$ & $S_\alpha$ (full complex remnant) & $S_\alpha$ (no
remnant)
\\ \hline
0.0 MeV & $5/2^+$ & 2 & 3 & 0.36 & 0.15 \\
0.87 MeV & $1/2^+$ & 3 & 1 & 0.42 & 0.18 \\
3.055 MeV & $1/2^-$ & 4 & 0 & 0.81 & 0.32 \\
3.84 MeV & $5/2^-$ & 3 & 2 & 0.64 & 0.28 \\
4.55 MeV & $3/2^-$ & 3 & 2 & 0.90 & 0.38 \\
6.356 MeV & $1/2^+$ & 4 & 1 & 0.49 & 0.24 \\ \hline
\end{tabular}
\end{table}

Figure~\ref{f1} shows that we obtain reasonably good descriptions of the data,
with the exception of that for transfers to the 3.84 MeV $5/2^-$ and 4.55 MeV
$3/2^-$ states at forward angles. Unlike the original analysis of Kubono {\sl et
al.\/} \cite{kubono}, we obtain a reasonable description of the 0.0 MeV $5/2^+$
state; as the $\alpha$ particle is in an $L = 3$ state relative to the $^{13}$C
core the DWBA angular distribution should be relatively structureless, like the
data. As we obtain such an angular distribution (both with FRESCO and DWUCK5),
in contrast to the structured angular distribution for this state shown in
Ref.~\cite{kubono}, we are forced to conclude that these authors have made an
error in their calculations for this state, either through plotting the wrong
curve or by using the wrong $L$ value.

It will be noted from Table \ref{t1} that if one adopts the procedure of Kubono
{\sl et al.\/} \cite{kubono} of re-normalizing all the $S_\alpha$ values with the
scaling factor that gives a value of $S_\alpha = 0.25$ for the 3.055 MeV $1/2^-$
state, derived from a microscopic cluster calculation \cite{furutani}, one
obtains from our DWBA results an $S_\alpha$ value for the 6.356 MeV $1/2^+$
state of 0.15 or 0.19, depending on whether the remnant term is included in the
calculations or not. Either value is over an order of magnitude greater than
that ($S_\alpha = 0.011$) found by Kubono {\sl et al.\/} \cite{kubono}. As
we have used the same potentials in our DWBA calculations as those used in
Ref.~\cite{kubono}, we are unable to account for this discrepancy in our results
for $S_\alpha$ compared to theirs. Note that the use of the remnant term cannot
account for the discrepancy (it is not clear from their paper whether Kubono
{\sl et al.\/} \cite{kubono} included a remnant term in their calculations)
as is apparent from Table \ref{t1}. The only uncertainty is in the values used for
the number of nodes $N$ in the $\alpha$ + $^{13}$C wave functions used by Kubono
{\sl et al.\/}, as these are not given in their original paper.

There is some ambiguity in the choice of $N$, depending on the structure assumed
for the state in question. In deriving the $N$ values for the 3.055 MeV $1/2^-$,
3.84 MeV $5/2^-$ and 4.55 MeV $3/2^-$  states given in Table \ref{t1} we have
assumed a 2p-1h configuration for these states. However, it has been suggested
that these states are of 4p-3h character \cite{bethge}, and if one assumes this
structure one obtains $N$ values of 5, 4 and 4 for the 3.055 MeV $1/2^-$, 3.84
MeV $5/2^-$ and 4.55 MeV $3/2^-$  states, respectively. These larger $N$ values
lead to $S_\alpha$ values of 0.30, 0.25 and 0.36, respectively, which would
yield a normalised $S_\alpha$ value of 0.41 for the 6.356 MeV $1/2^+$ state. In
reality, these states are probably a mixture of 2p-1h  and 4p-3h configurations
\cite{furutani,clark}; clearly, differences in the assumed structure of the
3.055 MeV $1/2^-$ state cannot account for the discrepancy in \emph{normalised}
$S_\alpha$ values between the current work and that of Kubono {\sl et al.\/}
\cite{kubono}.

For the 6.356 MeV $1/2^+$ state itself, the value of $N$ is more clearly
defined; any physically reasonable choice of structure for this state results in
a value of $N = 4$, the value we have adopted. Thus, we find that the smallest
value of $S_\alpha$ for this state consistent with the
$^{13}$C($^6$Li,$d$)$^{17}$O transfer data is 0.15 (after applying the
normalisation procedure discussed above) and that it could be as high as 0.41,
depending on the structure assumed for the 3.055 MeV $1/2^-$ state, used to
obtain the normalisation factor. It should be emphasised that there is no solid
justification for the normalisation procedure used in Ref.~\cite{kubono} to
obtain spectroscopic factors for \emph{all} states based on the $\alpha$-cluster
structure of the 3.055 MeV $1/2^-$ state. Furthermore, this procedure is
model-dependent as it relies strongly on the accuracy of the calculated
$S_\alpha$ of Furutani {\sl et al.\/} \cite{furutani} for the 3.055 MeV $1/2^-$
state.

We are thus forced to conclude that the original DWBA analysis of Kubono {\sl et
al.\/} \cite{kubono} is flawed, and that the $S_\alpha$ value obtained by them for
the 6.356 MeV $1/2^+$ state is much too small. Our DWBA calculations using the
same optical potential parameters as those used by Kubono {\sl et al.\/} show that
the $^{13}$C($^6$Li,$d$)$^{17}$O transfer data measured by them are consistent
with an (un-renormalised) $S_\alpha\approx 0.49$ for the 6.356 MeV $1/2^+$ state of
$^{17}$O, well within the range used in astrophysical calculations
($S_\alpha\approx 0.3-0.7$).

\begin{figure}
\begin{center}
\psfig{figure=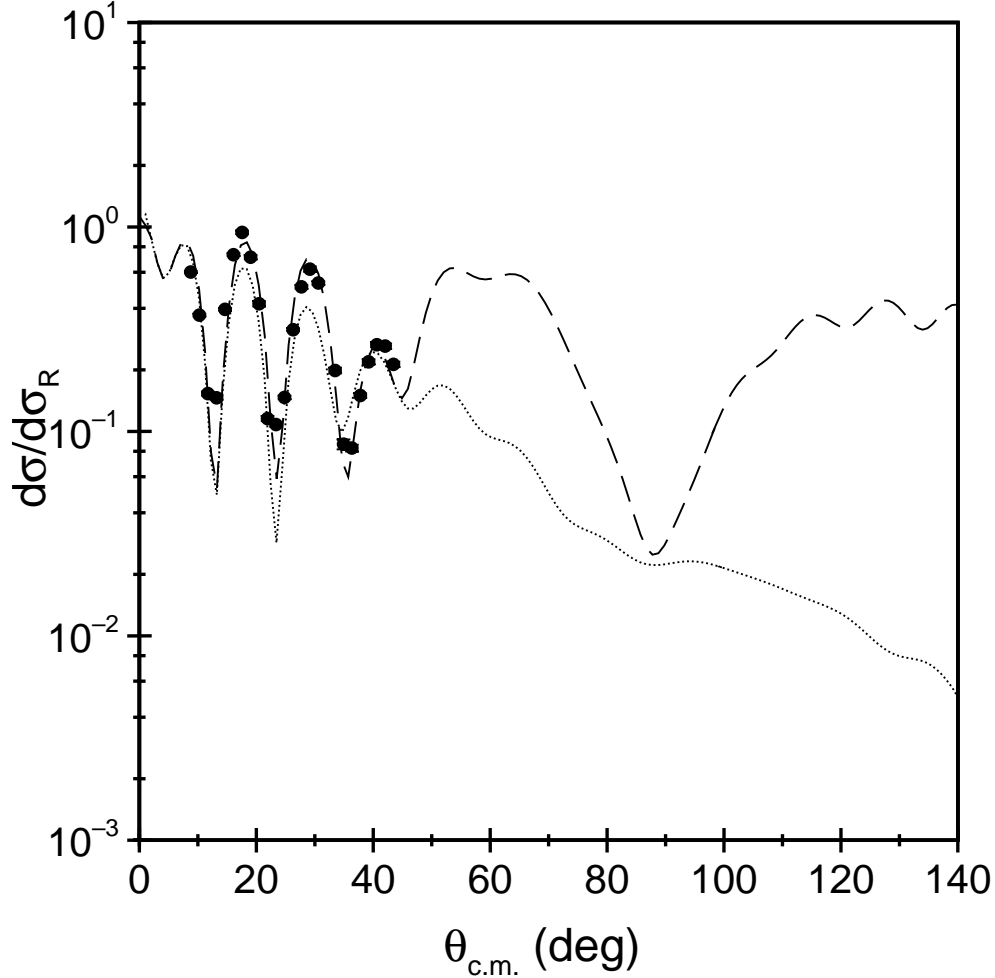,clip=,width=13.0cm}
\end{center}
\caption{\label{f2}Optical model calculations compared to the $^6$Li+$^{13}$C
elastic scattering data. The dashed curve denotes the predicted elastic
scattering angular distribution using the optical model potential set 1 of
Kubono {\sl et al.\/} \cite{kubono}. The dotted curve denotes the prediction of the
present Folding + DPP potential.}
\end{figure}

We now discuss a subtle but quite important issue: the sensitivity of the DWBA
calculation to the choice of optical potential for the entrance channel. This
potential defines the door-way state of the transfer reaction and it should
be determined as accurately as possible. The $^6$Li+$^{13}$C elastic
scattering was measured by Kubono {\sl et al.\/} at forward angles only
\cite{kubono}, which gives rise to some ambiguity in the choice of the optical
potential. In Fig.~\ref{f2} we plot the optical model fit to the $^6$Li+$^{13}$C
elastic scattering data using their parameter set 1. One can see that for large
angles the predicted cross section exhibits unphysical behaviour compared to
data for $^6$Li + $^{12}$C elastic scattering at the same $^6$Li incident energy
which extend to backward angles \cite{bingham}.

The $^6$Li+$^{12}$C elastic scattering has been extensively studied in the past,
and a very accurate systematics for the (energy dependent) optical potential has
been established in a folding model analysis of these data over 
a wide range of incident energies \cite{Kho95}. An important
result of this study is that the so-called dynamic polarisation potential (DPP)
caused by $^6$Li breakup has a strong \emph{repulsive} contribution to the real
potential. This DPP contribution can be rather well represented by a
surface Woods-Saxon (WS) potential. We have, therefore, performed an additional
optical model analysis of the $^6$Li+$^{13}$C elastic scattering data at $E_{\rm
lab}=60$ MeV using an entrance channel optical potential of the following form: 
\begin{equation}
U(R)=V_{\rm Fold}(R)+\Delta V(R)+iW(R), \label{e2}
\end{equation}
\begin{equation}
{\rm where}\ \Delta V(R)=4Va_V\frac{d}{dR}\left[1+\exp
\left(\frac{R-R_V}{a_V}\right)\right]^{-1} \label{e3}
\end{equation}
\begin{equation}
{\rm and}\ W(R)=-W\left[1+\exp \left(\frac{R-R_W}{a_W}\right)\right]^{-1}.
\label{e4}
\end{equation}
The DPP and volume WS imaginary potential parameters obtained (see Table~\ref{t2})
are quite close in shape to those determined from the Folding model analysis
\cite{Kho95} of the $^6$Li+$^{12}$C elastic scattering data at the same
incident $^6$Li energy but over a much wider angular range.

\begin{table}\small
\caption{\small Parameters of the real DPP and volume WS imaginary potentials
(\ref{e3})-(\ref{e4}). The negative value for $V$ indicates a repulsive
potential. Radius parameters are given as: $R_x = r_x \times 13^{1/3}$.}
\label{t2}
\begin{tabular}{c c c c c c} \hline
$V$ (MeV) & $r_V$ (fm) & $a_V$ (fm) & $W$ (MeV) & $r_W$ (fm) & $a_W$ (fm) \\
\hline -18.0403& 1.5079 & 0.6823 & 37.2152 & 1.3770 & 1.0150 \\ \hline
\end{tabular}
\end{table}

The Folding potential was calculated using the CDM3Y6 interaction (whose
density-dependent parameters were fine-tuned to reproduce the bulk properties of
cold nuclear matter \cite{khoa}) and the ground-state densities of $^6$Li and
$^{13}$C obtained in the independent-particle model by Satchler and
Love \cite{sat1} and Satchler \cite{sat2}, respectively. The predicted elastic
scattering obtained using this potential is shown as the dotted curve in
Fig.~\ref{f2}. Note that the volume integral of the real optical potential per
interacting nucleon pair $J_V$ (an important key to distinguish discrete
potential families) is around -410 and -329 MeV~fm$^3$ for the set 1 potential
taken from Ref.~\cite{kubono} and our Folding + DPP potential, respectively. The
empirical energy dependence of $J_V$ is well-known from the global optical model
analysis of light heavy-ion elastic data \cite{Bra97a}, and one can find from
this systematics that $J_V$ should be around -330 MeV\ fm$^3$ for an energy of
10 MeV/nucleon (see Fig.~6.7 in Ref.~\cite{Bra97a}) which is in perfect
agreement with our Folding + DPP calculation. We conclude, therefore, that the
Folding + DPP potential should be more appropriate for the entrance channel optical
potential. While both Kubono {\sl et al.\/}'s set 1 and our Folding + DPP
potential produce similar elastic scattering cross sections at the forward angles for which
data are available, they differ considerably at larger angles, and a future
measurement of the $^6$Li+$^{13}$C elastic scattering at larger angles would be
extremely helpful in determining a realistic entrance channel optical potential.

\begin{figure}
\begin{center}
\psfig{figure=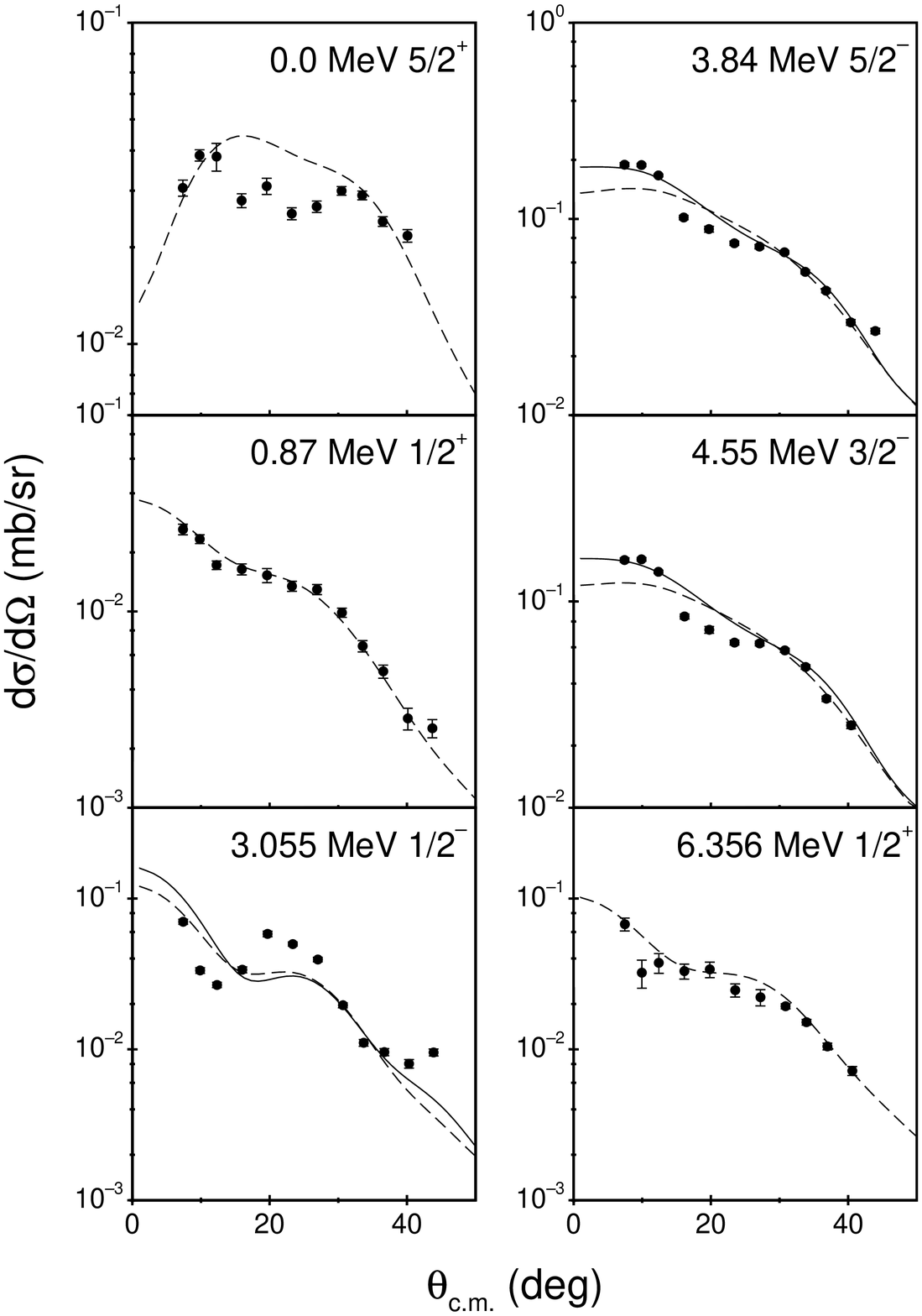,clip=,width=13.0cm}
\end{center}
\caption{\label{f3}DWBA calculations using the Folding + DPP optical potential
for the $^6$Li+$^{13}$C system compared to the data. For the 3.055 MeV $1/2^-$, 3.84
MeV $5/2^-$ and 4.55 MeV $3/2^-$ states the dashed curves denote calculations
assuming a 2p-1h structure, while the solid curves denote calculations assuming
a 4p-3h structure. All calculations use the post form of the DWBA.}
\end{figure}

We show the results of the DWBA calculations using this new $^6$Li +
$^{13}$C optical potential in Fig.~\ref{f3} and give the $S_\alpha$ values
obtained from this analysis in Table \ref{t4}. As can be seen from
Fig.~\ref{f3}, with the exception of the 3.055 MeV $1/2^-$ state, the agreement
between the DWBA calculations and the data is much better than that obtained
using the set 1 potential of Kubono {\sl et al}. This is particularly noticeable
for the two $1/2^+$ states. Table \ref{t4} also shows that the $S_\alpha$ values
obtained from this analysis are considerably smaller than those given in Table
\ref{t1}. The absolute $S_\alpha$ value obtained for the 6.356 MeV $1/2^+$ state
is around 0.36 in this case. If we adopt the (controversial) normalisation
procedure discussed above, we obtain $S_\alpha$ values of 0.3 or 0.5 for the
6.356 MeV $1/2^+$ state, depending on whether a 2p-1h or 4p-3h structure is
assumed for the 3.055 MeV $1/2^-$ state. All these estimates of the $S_\alpha$
value are over an order of magnitude larger than that ($S_\alpha\approx 0.011$)
determined in Ref.~\cite{kubono}.

The influence of the entrance channel optical potential on the $S_\alpha$ values
extracted from the DWBA analysis stresses again the importance of obtaining
elastic scattering data over a sufficiently wide angular range for light
heavy-ion systems such as that currently under consideration.

\section{Coupled reaction channels calculations}
In order to test the possible influence of multi-step transfer paths on the
$S_\alpha$ values we also carried out a limited coupled reaction channels (CRC)
analysis which included couplings between the ground state of $^{17}$O and its
$1/2_1^+$, $1/2_1^-$, $5/2_1^-$, $3/2_1^-$ and $1/2_2^+$ excited states as well
as the direct $\alpha$ transfer route. We did not consider multi-step paths
proceeding via the excited states of $^{13}$C as a physically meaningful
analysis including these paths is not possible without some prior estimate of
the necessary $\alpha$ strengths.

We used the Folding + DPP optical potential described in the previous section in
the entrance channel and parameter set 3 of Kubono {\sl et al.\/} \cite{kubono} in
the exit channel. The $^{17}$O Coulomb coupling strengths were determined from
the measured $B(E\lambda)$ values \cite{manley} and the nuclear coupling
potentials were obtained by deforming the $d$+$^{17}$O optical potential with
nuclear deformation lengths obtained from the corresponding $B(E\lambda)$ values
(assuming an $^{17}$O radius of $1.2 \times 17^{1/3}$ fm). As the $^{17}$O
coupling strengths are weak, it was not found necessary to alter the
$d$+$^{17}$O optical potential to compensate for the inclusion of these
couplings.

\begin{figure}
\begin{center}
\psfig{figure=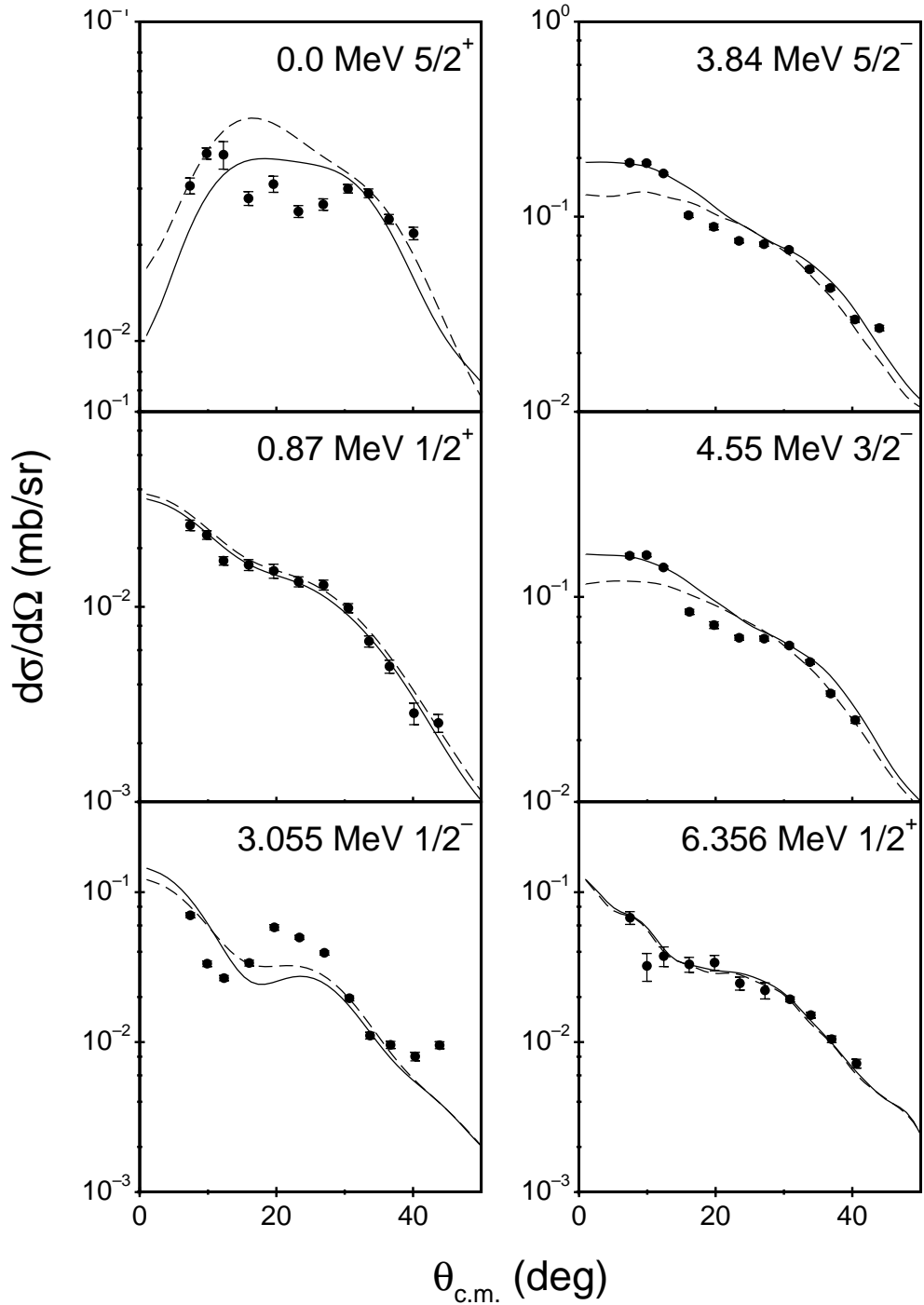,clip=,width=13.0cm}
\end{center}
\caption{\label{f4}CRC calculations using the Folding + DPP optical potential
for the $^6$Li+$^{13}$C system compared to the data. The dashed curves denote the
result of a calculation assuming a 2p-1h structure for the 3.055 MeV $1/2^-$,
3.84 MeV $5/2^-$ and 4.55 MeV $3/2^-$ states, while the solid curves denote the
result of a calculation assuming a 4p-3h structure for these states.}
\end{figure}

The results of the CRC analysis are shown in Fig.~\ref{f4}. The CRC description
of the elastic scattering is almost identical to that shown by the dotted curve in
Fig.~\ref{f2}. The $S_\alpha$ values extracted from this CRC analysis are given
in Table~\ref{t4} in the two columns labelled ``CRC(2p-1h)'' and ``CRC(4p-3h)'',
which denote the values obtained from the calculations assuming a 2p-1h and
4p-3h structure, respectively, for the $1/2_1^-$, $5/2_1^-$ and $3/2_1^-$
states.

\begin{table}\small
\caption{\small Values of $S_\alpha$ obtained from the DWBA and CRC calculations
using the Folding + DPP optical potential for $^6$Li+$^{13}$C system.}
\label{t4}
\begin{tabular}{c c c|c|c|c} \hline
  &  &   & DWBA       & CRC (2p-1h) & CRC (4p-3h) \\
State & $J^\pi$ & $N$ & $S_\alpha$ & $S_\alpha$ & $S_\alpha$ \\ \hline
0.0 MeV & $5/2^+$ & 2 & 0.15 & 0.12 & 0.15 \\
0.87 MeV & $1/2^+$ & 3 & 0.17 & 0.17 & 0.17 \\
3.055 MeV & $1/2^-$ & 4 & 0.30 & 0.30 &  \\
          &        & 5 & 0.18 &  & 0.18 \\
3.84 MeV & $5/2^-$ & 3 & 0.34 & 0.34  & \\
          &       & 4 & 0.19 &  & 0.19 \\
4.55 MeV & $3/2^-$ & 3 & 0.48 & 0.49  & \\
          &       & 4 & 0.27 &   & 0.27 \\
6.356 MeV & $1/2^+$ & 4 & 0.36 & 0.40 & 0.40 \\ \hline
\end{tabular}
\end{table}

As Fig.~\ref{f4} shows, with the exception of the transfer to the $5/2^+$ ground
state, the shapes of the predicted angular distributions are unchanged by the
inclusion of couplings between the ground state and the excited states of
$^{17}$O. The $S_\alpha$ values are also either unchanged or, for the $5/2_1^+$
and $1/2_2^+$ states, very slightly altered by these extra couplings. The
changes in the $S_\alpha$ values for the $5/2_1^+$ and $1/2_2^+$ states are
smaller than those produced by, for example, the use of the Folding + DPP
optical potential for the $^6$Li+$^{13}$C system compared to Kubono {\sl et al.\/}'s
set 1 potential or by the inclusion of the full complex remnant term compared to
a calculation with no remnant term. Therefore, we may conclude that couplings
within $^{17}$O will have a negligible effect on the $S_\alpha$ values extracted
from an analysis of the $^{13}$C($^6$Li,$d$)$^{17}$O transfer reaction. For the
crucial 6.356 MeV $1/2^+$ state, an $S_\alpha$ value around 0.4 is given by both
DWBA and CRC calculations which is fully compatible with those used earlier
($S_\alpha\approx 0.3-0.7$) in the $s$-process model calculations.

However, multi-step effects can be important and lead to anomalous population of
non $\alpha$-cluster states in both the ($^6$Li,$d$) and ($^7$Li,$t$) reactions,
as was noted by Debevec {\sl et al.\/} \cite{debevec} for the $^{12}$C($^6$Li,$d$)$^{16}$O
reaction. They found that for 32 MeV incident $^6$Li the 10.35 and 11.1 MeV $4^+$
states are populated with relative strengths of $\sigma(11.1)/\sigma(10.35) \approx
0.5$ despite the fact that their $\alpha$-particle widths are in the ratio
$\Gamma_\alpha(11.1)/\Gamma_\alpha(10.35) \approx 0.01$. Kemper and Ophel \cite{kemper}
have shown that this anomalous population of the 11.1 MeV $4^+$ state is non-statistical
in origin and must therefore be due to multi-step transfer paths. This problem is
not so marked for the $^{12}$C($^7$Li,$t$)$^{16}$O reaction and appears to become
less important as the incident $^6$Li or $^7$Li energy increases. Nevertheless, it
is an indication that multi-step paths can be important and must be considered
carefully if one wishes to extract meaningful $S_\alpha$ values from the analysis of
$\alpha$ transfer reactions.

\section{Reduced $\alpha$ widths}
One may also derive the reduced $\alpha$ width, $\gamma_\alpha^2$, of a state
from a DWBA analysis. The reduced $\alpha$ width is related to the
spectroscopic factor and the $\alpha+^{13}$C bound state radial wave function used in
the DWBA calculations as follows
\begin{equation}
\gamma_\alpha^2 = S_\alpha \frac{\hbar^2 R}{2 \mu_\alpha} \left| u_L(R)\right|^2
\end{equation}
where $\mu_\alpha$ is the $\alpha+^{13}$C reduced mass, $R$ is the channel
radius and $u_L(R)$ is the $\alpha+^{13}$C bound state radial wave function, normalised
such that $\int_0^\infty u_L^2(r) r^2 dr = 1.0$. We take $R = 5.5$ fm as our
channel radius, the value used by Kubono {\sl et al.\/} \cite{kubono} and Furutani
{\sl et al.\/} \cite{furutani}. Such a choice for the channel radius is reasonably
compatible with the `strong absorption' radius of the $\alpha+^{13}$C system of
around 5.2--5.3 fm. The reduced $\alpha$ width is often scaled by the Wigner
limit, defined as
\begin{equation}
\gamma_W^2(R) = \frac{3 \hbar^2}{2 \mu_\alpha^2 R^2},
\end{equation}
to obtain the dimensionless reduced $\alpha$ width, $\theta_\alpha^2(R)$
\cite{becchetti}. It should be noted that Kubono {\sl et al.\/} \cite{kubono} have
made assumptions about the $\alpha+^{13}$C bound state wave function that are
equivalent to having $S_\alpha=\theta_\alpha^2$, which is not always the case;
we have not made this assumption here.

In Table~\ref{t3} we give values for $\gamma_\alpha^2$ and $\theta_\alpha^2$
calculated using the $S_\alpha$ values of Tables~\ref{t1} and \ref{t4}. Note
that we do not use the normalised $S_\alpha$ values in calculating
$\gamma_\alpha^2$ and $\theta_\alpha^2$, as it is uncertain what the effect of
the scaling process on the $\alpha+^{13}$C bound state wave function should be.

\begin{table}\small
\caption{\small $\theta_\alpha^2$ and $\gamma_\alpha^2$ values obtained from
DWBA analyses using the $^6$Li+$^{13}$C optical potentials of Kubono {\sl et
al.\/} \cite{kubono} (Set 1) and the present work (Folding + DPP) at a channel
radius $R = 5.5$ fm. Also shown are values for $\theta_\alpha^2(5.5)$ from the
microscopic cluster-model calculation of Furutani {\sl et al.\/} \cite{furutani}.}
\label{t3}
\begin{tabular}{c c c|c c|c c|c} \hline
 & & & \multicolumn{2}{c}{Set 1} & \multicolumn{2}{c}{Folding + DPP} &
 Ref.~\cite{furutani} \\
State & $J^\pi$ & $N$ & $\theta_\alpha^2$ & $\gamma_\alpha^2$ (keV) &
$\theta_\alpha^2$ & $\gamma_\alpha^2$ (keV) &$\theta_\alpha^2$ \\
\hline
0.0 MeV & $5/2^+$ & 2 & 0.209 & 142 & 0.087 & 59.1 & \\
0.87 MeV & $1/2^+$ & 3 & 0.331 & 224 & 0.134 & 90.8 & \\
3.055 MeV & $1/2^-$ & 4 & 0.723 & 490 & 0.268 & 182 & 0.084\\
             &     & 5 & 0.254 & 172 & 0.153 & 104 & \\
3.84 MeV & $5/2^-$ & 3 & 0.554 & 375 & 0.294 & 199 & 0.23\\
             &    & 4 & 0.216 & 146 & 0.164 & 111 & \\
4.55 MeV & $3/2^-$ & 3 & 0.760 & 515 & 0.406 & 275 & 0.16\\
             &    & 4 & 0.288 & 195 & 0.216 & 146 &\\
6.356 MeV & $1/2^+$ & 4 & 0.272 & 184 & 0.200 & 136 & \\ \hline
\end{tabular}
\end{table}
We have checked our procedures against those of Becchetti {\sl et al.}
\cite{becchetti} for the much more widely studied $^{12}$C($^7$Li,$t$)$^{16}$O
reaction and obtain good agreement with their values for $\gamma_\alpha^2$ for
states in $^{16}$O using the same $\alpha+^{12}$C wave functions and $S_\alpha$
values. It should be emphasised that we obtain values of $\gamma_\alpha^2$ for
the crucial 6.356 MeV $1/2^+$ state considerably greater than the 7.4 keV value
given in Ref.~\cite{kubono}.

We also give in Table~\ref{t3} $\theta_\alpha^2$ values for the 3.055 MeV
$1/2^-$, 3.84 MeV $5/2^-$ and 4.55 MeV $3/2^-$ states from the microscopic
cluster-model calculation of Furutani {\sl et al.\/} \cite{furutani}. The
calculations using Kubono {\sl et al.\/}'s set 1 for the $^6$Li + $^{13}$C optical
potential yield $\theta_\alpha^2$ values for the 3.055 MeV
$1/2^-$ state that are much smaller than the cluster model value, while for
the 3.84 MeV $5/2^-$ and 4.55 MeV $3/2^-$ states our $\theta_\alpha^2$ values
assuming a 4p-3h structure for these states are reasonably close to Furutani
{\em et al.\/}'s cluster model results. For the calculations using our Folding + DPP optical potential
for the $^6$Li+$^{13}$C system, the agreement is better and the cluster-model
$\theta_\alpha^2$ values agree reasonably well with our values assuming a 4p-3h
structure for these three states. This agreement with the calculation of
Furutani {\sl et al.\/} \cite{furutani} shows that the $\theta_\alpha^2$ values
that we have extracted from our DWBA calculations are at least plausible.
Furthermore, it supports the suggestion of Bethge {\sl et al.} \cite{bethge}
that these states (the 3.055 MeV $1/2^-$, 3.84 MeV $5/2^-$ and 4.55 MeV $3/2^-$)
are of predominantly 4p-3h character.

\section{Conclusions}
A recent DWBA analysis of new 60 MeV $^{13}$C($^6$Li,$d$)$^{17}$O
transfer data \cite{kubono} found a very small $\alpha$-spectroscopic factor
($S_\alpha\approx 0.011$) for the 6.356 MeV $1/2^+$ state. 
The authors of Ref.\ \cite{kubono} concluded that there is no large enhancement of the
$^{13}$C($\alpha$,$n$)$^{16}$O reaction rate at energies of astrophysical
interest due to this sub-threshold state based on the small reduced $\alpha$
width they obtained, contrary to previous suggestions
\cite{drotleff,hale}. Since such a result poses serious questions about the
$^{13}$C($\alpha$,$n$)$^{16}$O reaction as a neutron source for the $s$ process,
we have performed DWBA and CRC analyses of their data using a realistic
choice for the entrance channel optical potential. We concluded that 
the transfer data are consistent
with a much larger $S_\alpha$ value for the 6.356 MeV $1/2^+$ state and
consequently a much larger $\gamma_\alpha^2$ value. Thus, the transfer data
are compatible with a large contribution from this state to the
$^{13}$C($\alpha$,$n$)$^{16}$O reaction rate at very low energies.

We have also shown that uncertainties in the assumed $\alpha+^{13}$C structure
of the $^{17}$O states can have significant effects on the extracted
spectroscopic factors and reduced $\alpha$ widths. While this problem does not
directly affect the 6.356 MeV $1/2^+$ state, one needs to be aware of it. As \emph{relative} $S_\alpha$ 
factors are reasonably well determined one may use the normalisation procedure employed by
Kubono {\sl et al.\/} to scale the DWBA results to a calculated value for a state that
is believed to have significant $\alpha$ clustering structure in order to 
obtain absolute $S_\alpha$ values. However, such a method is highly questionable as the fragmentation of the
$\alpha$-cluster states over the low-energy region of $^{17}$O excitations is
unknown. Moreover, there are difficulties in this method as it relies on the
accuracy of the structure calculation used for the normalisation, plus it is not
clear what effect this process should have on the $\alpha$ + core wave function
if one wishes to extract $\gamma_\alpha^2$ values.

We also noted the effect of the entrance channel optical potential on the
$S_\alpha$ values extracted from the DWBA analysis, which was considerable. This
demonstrates the need to have a realistic choice for the optical potential in the
entrance channel. It is, therefore, necessary to obtain elastic
scattering data over a large angular range (possibly much larger than the
angular range of the transfer data themselves) if one wishes to use this type of
indirect approach to obtain reliable estimates of astrophysical reaction rates.

A limited coupled reaction channels study found that couplings between the
ground and excited states of $^{17}$O had a negligible effect on the $S_\alpha$
values. Other multi-step transfer paths, in particular, those proceeding via the
excited states of $^{13}$C, were not investigated due to the need for some prior
estimate of the necessary $\alpha$ strengths. Without such estimates one merely
introduces extra adjustable parameters that need to be determined from the same
data set. 

In summary, we have highlighted some of the pitfalls of DWBA analyses of
$\alpha$ transfer reactions and their use to extract spectroscopic factors. We
have not considered in detail the more fundamental question of whether such
transfer reactions may be adequately described by a simple direct $\alpha$
transfer, as is assumed in the DWBA. If multi-step paths contribute
significantly to the $\alpha$ transfer strength the $S_\alpha$ values extracted
from such a DWBA analysis will not be reliable, regardless of how carefully it
is carried out. However, an extensive investigation of these multi-step transfer
paths still cannot be carried out in a meaningful way for the
$^{13}$C($^6$Li,$d$)$^{17}$O reaction due to the lack of prior estimates of the
required $\alpha$ strengths for transfers proceeding via the excited states of
$^{13}$C. Thus, while the indirect approach to astrophysical reaction rates can
be a valuable tool one needs to be aware of possible uncontrolled parameters
contained in the analysis.

\section*{ACKNOWLEDGEMENTS}
The authors thank Prof.\ S. Kubono for providing them with the data in tabular
form and Prof.\ D. Robson for many helpful discussions. One of the authors
(N.K.) would like to thank the Institute for Nuclear Science and Technique,
Hanoi, for hospitality during the period in which this work was initiated. The
present research was supported, in part, by the Natural Science Council of
Vietnam, the U.S. National Science Foundation and the State of Florida.

\end{document}